# Anomalous wave reflection at the interface of two strongly nonlinear granular media

V. F. Nesterenko[1,2*], C. Daraio[1], E.B. Herbold[2], S. Jin[1,2]

*[1]Materials Science and Engineering Program*

*[2]Department of Mechanical and Aerospace Engineering*

*University of California at San Diego, La Jolla CA 92093-0418 USA*

ABSTRACT

Granular materials exhibit a strongly nonlinear behaviour affecting the propagation of information in the medium. Dynamically self-organized strongly nonlinear solitary waves are the main information carriers in granular chains. Here we report the first experimental observation of the dramatic change of reflectivity from the interface of two granular media triggered by a noncontact magnetically induced initial precompression. It may be appropriate to name this phenomenon the "acoustic diode" effect. Based on numerical simulations, we explain this effect by the high gradient of particle velocity near the interface.

Strongly nonlinear granular chains are "sonic vacuum" (SV) type systems that support a new type of solitary waves [1,2,3]. These solitary waves are qualitatively different from the well known weakly nonlinear solitary waves of the Korteweg-de Vries equation [4,5] which were first discovered experimentally by Russel [6,7] in 1834. The concept of SV was proposed to emphasize the uniqueness of the types of materials which do not support sound waves without initial prestress [3,8]. One of the main features of strongly nonlinear solitary waves is that their speed is strongly influenced by interrelated potential and kinetic energies [3]. A granular chain with particles interacting according to Hertz law is just one of the examples of strongly nonlinear behaviour. Different groups investigated numerically and experimentally the properties of these waves [1,2,9-14] and found a good agreement with the theoretical predictions based on a long wave approximation [1,3]. Nonlinear dynamic properties can be extended to other designed metamaterials including the propagation of electrical or other types of signals. Interesting applications of this new area of wave dynamics have been proposed, for example for creation of nanodroplets [15].

One of the intriguing properties of these materials is the reflection of the solitary waves at the interface of two SV materials or from a wall [3,8,14,18,19]. Based on this behaviour, the novel concept of impulse trapping inside a protecting granular laminar layer has been proposed [16,17].

In this Letter we present first experimental and numerical observation of strongly nonlinear wave interaction with the interface of two SV-type systems resulting in anomalous reflected compression and transmitted rarefaction waves when the magnetically induced prestress is applied.

We propose a new method of noncontact precompression based on the magnetic interaction of the first magnetic particle in the chain with a Nd-Fe-B ring magnet placed outside the chain [20]. The magnetic force is practically independent of the motion of the magnetic particle. It allows a direct control of the boundary conditions, unlike the precompression induced by attaching loads with wires [3] or the contact precompression by an outside frame [9].

In experiments we placed a chain of 20 nonmagnetic stainless steel (316) particles (plus a magnetic particle on the top) above 21 PTFE (polytetrafluoroethylene) beads. Piezogauges were placed inside particles measuring the averaged compression forces between the two corresponding contacts [21].

The experimental and numerical results without magnetic precompression are presented in Fig. 1(a) and (b). A single solitary wave (I) was excited in the system. After the interaction with the interface this solitary wave excites a few pulses (T) in the PTFE chain. No compression reflected wave is detected in the stainless steel chain in experiments and in numerical calculations (Fig. 1(a) and (b)) as in [3,8,18].

The mechanism explaining the practically complete energy transfer into the PTFE chain and the absence of a reflected compression wave is illustrated by the numerical results for the particles displacements in Fig. 1(c). It is evident that a series of gaps (see cartoon in Fig. 1(d)) opens between the stainless steel particles in the vicinity of the interface. Therefore a “fracture wave” is propagating into the stainless steel chain similar to the one originating when a solitary wave arrives at the free surface [11] due to high gradients of particle velocities in the wave. Gaps opening in the granular media is observed in 2-D numerical simulations of the transmission of the static force [22,23] and is related to the subharmonics and noise excitation in the transmission of the acoustic wave [24].

The gaps opening and closing introduce an entirely new time scale in the system which is determined by the size of the gaps and particle velocities instead of the time of flight determined by the size of the system and signals speed.

The strongly nonlinear character of the particles interaction results in a high gradient of particle velocity in the incident wave translating into the high gradient of velocity near the interface with the last stainless steel particle absorbing the main part of the energy. The zero tensile strength in a granular matter ensures the uni-directional energy transfer to the PTFE chain without sending any tensile wave back to the stainless steel chain. We can see that the observed behaviour is due to a double nonlinearity: combination of a strongly nonlinear compression part of the interacting force and a zero tensile strength of the system.

At the moment when the process of energy transmission into the PTFE chain is practically completed only a very small portion of the kinetic energy of the impactor (about 0.16%) is reflected and the second steel particle is moving back with velocity 0.018 m/s. Without the gravitational precompression the second steel particle moves back much later with a velocity significantly smaller (0.0009 m/s) than the one observed in the previous case. This indicates that the reflected energy can be increased with the initial precompression.

The sequence of pulses in the PTFE chain is generated by the decelerating interfacial stainless steel particle which is demonstrated by the kinks of decreasing amplitude on the displacement curve for the first PTFE particle in Fig. 1(c).

The application of the magnetically induced precompression (2.38 N) resulted in a completely different reflection of strongly nonlinear pulse from the interface (compare Fig. 1(a), Fig. 1(b) and Fig. 2(a), Fig. 2(b)).

As in the previous case (Fig. 1(a) and (b)), a single pulse is propagating into the steel chain followed by an oscillatory nonstationary rarefaction wave (Fig. 2(a) and (b)). In this particular combination, the acoustic impedance of the stainless steel chain is about one order of magnitude higher than that of the PTFE chain which should result in the linear approach only in a rarefaction wave propagating back into stainless steel chain from the interface. The rarefaction wave is indeed noticeable from the coordinated change of sign of the slope in the displacements curves starting from the $4^{th}$ stainless steel particle adjacent to the interface (Fig. 2(c)). But additionally to the expected reflected rarefaction wave we observed experimentally and numerically anomalous reflected compression waves (Fig. 2(a) and (b))! The leading reflected pulse has an amplitude of about one half of the amplitude of the incident wave. At the same time the leading transmitted compression pulse in PTFE chain is followed by unexpected rarefaction pulses (Fig. 2(a) and (b)).

A peculiar characteristic of this reflected compression wave is related to its delayed time of arrival at the gauges inside stainless steel particles in experiments and calculations (Fig. 2(a) and (b)).

When the reflected compression wave is formed, at 500 microseconds after the impact, the energy transferred to the PTFE chain is about 86% of the combined kinetic energy of the striker and the energy supplied by the magnetic force at the early stage of the motion of the first particle (about 25 microseconds). This contrasts with the previous case where almost all kinetic energy of the striker (over 99%) was transferred into the PTFE chain. It may be appropriate to name this dramatic change of reflectivity triggered by the initial precompression (zero and 14% reflected energy correspondingly) the “acoustic diode” effect.

The reflected compression waves were formed mainly due to the rebounding motion of the first stainless steel particle due to the resistance of the PTFE beads. The particles in the vicinity of interface reorganize into a state close to the original precompressed state (Fig. 2(c)).

As in the previous case (Fig. 1(c)), the stainless steel interfacial particle serves as the main energy transformer from the stainless steel chain into the PTFE chain (Fig. 2(c)). Even with the preloading, which would intuitively prevent the formation of gaps, their characteristic opening and closing also dominate the process of wave reflection in this case (Fig. 2(c) and (d)).

Again, a “fracture-wave” following the rarefaction wave is propagating from the interface back into the stainless steel chain. In this case of stronger precompression the gaps are closed rather quickly allowing the formation of reflected compression pulses with an amplitude of about one half of the incident wave. The gaps opening and closures again introduce a new time scale which is determined by the size of the gaps and particle velocities. This new time scale is much shorter than in the previous case (Fig. 1(c)) due to the significantly smaller size of the opened gaps and larger velocities of the particles moving into the gaps (compare Figs. 2(c) and 3(c)).

Such behaviour is a consequence of the strongly nonlinear compressive interaction combined with the zero tensile strength of the system. Initial precompression in a counterintuitive manner triggers the generation of a reflected compression wave and does not suppress the process of gaps opening and closures but instead makes it faster. This indicates that the response of the interface between two SVs can be qualitatively tuned by the applied static preloading.

In our experiments, gap opening between second and interfacial stainless steel particles has a threshold on the amplitude of the incident wave equal about 3 N being larger than preload at the interface 2.473 N. The arrival of the anomalous reflected compression wave was detected in numerical calculations even at this amplitude with a 36 microsecond delay. This delay is due to a rather long duration of the interaction between the interfacial stainless steel particle and the first PTFE bead caused by the lower elastic stiffness of the contact. This means that the anomalous reflected compression wave is not caused by gaps opening but by the changing direction of the velocity of interfacial stainless steel particle. The gap opening significantly increases the delay time of arrival of this compression wave at the corresponding gauges placed inside stainless steel particles.

The peculiar motion of only two stainless steel particles adjacent to the interface (discrete level phenomenon) results in the pattern of reflected compression solitary waves which can be described in terms of continuum approach.

To clarify the process of the reflection further, the interaction of a solitary wave (Fig. 3(a)) with a similar interface including a larger number of particles was simulated (Fig. 3). After the interaction, a rarefaction wave was formed close to the interface in the stainless steel chain and it was followed by a compression pulse and an oscillatory tail (Fig. 3(b) and (c)). When the reflected signal reaches the 200$^{th}$ particle from the interface the anomalous compression solitary wave almost surpasses the rarefaction wave becoming the leading pulse dominating the reflection process (Fig. 3(b)). The behaviour of the displacements of the particles adjacent to the interface is very close to their behaviour in our experimental set-up (compare Figs. 2(c) and 3(c)). The leading transmitted compression pulse in the PTFE chain is followed by an unexpected rarefaction wave and an oscillatory tail (Fig. 3(b) and (c)).

At the moment depicted in Fig. 3(b) the energy exchange between the two chains is practically finished, the interfacial velocity is close to zero and the force to the initial prestress. Nonetheless the leading reflected and transmitted compression pulses continue to propagate through the chains, followed by evolving oscillatory tails.

In experiments and numerical calculations we did not observe a qualitative change of the reflectivity under the applied precompression when the wave approached the interface from the PTFE side.

In summary we observed a strong sensitivity on the initial precompression of the reflected and transmitted energy from the interface of the two granular media. This phenomenon can be named as “acoustic diode” effect. It can be employed for designing tunable information transportation lines with the unique possibility to manipulate the signals delay and reflection at will, and decompositions/scrambling of security-related information. It can also be used for identification of such interfaces (i.e. geological multilayer structure consisting of dissimilar granular materials) and for optimization of shock protection layers composed from uniquely combined composite granular media containing layers with different particle sizes (masses) and elastic constants.

The authors wish to acknowledge the support of this work by the US NSF (Grant No. DCMS03013220).

* Electronic address: vnesterenko@ucsd.edu

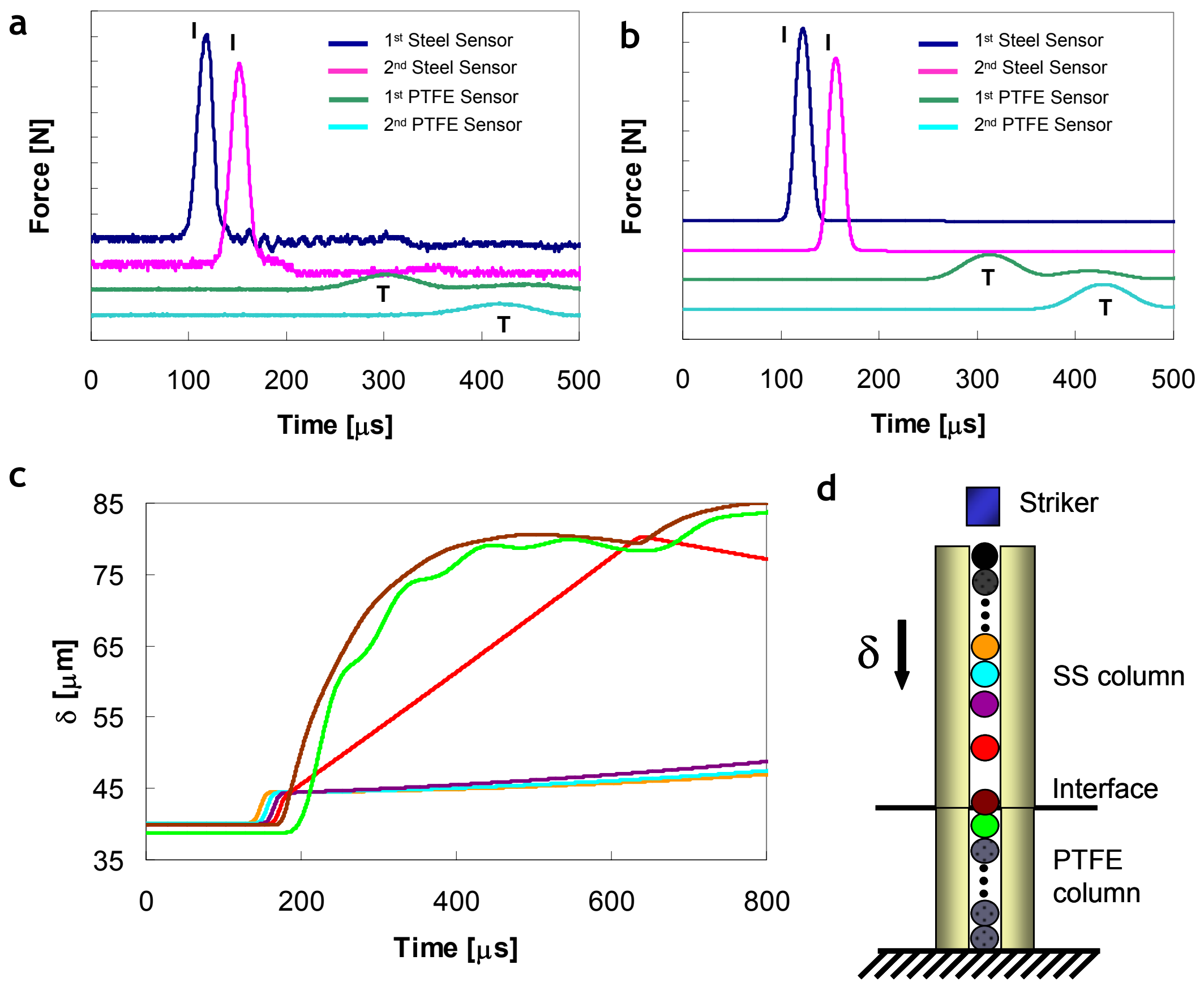

FIG. 1. Pulse reflections from the interface of the two SVs composed by 20 stainless steel particles (a magnetic steel particle was placed on the top, no magnetically induced precompression) above 21 PTFE beads. The impact was by alumina cylinder with mass 0.47 g, velocity 0.44 m/s. Dark blue lines in (a),(b) are the incident waves (I) detected by the sensor in the 8$^{th}$ particle and pink is the signal from the 4$^{th}$ particle above the interface, green is the transmitted signal (T) from the sensor in the 4$^{th}$ and light blue is the signal detected in the 8$^{th}$ particle below the interface. No wave reflected from interface into stainless steel chain was detected. The curves represent the averaged forces on the contacts of each particle. (a) Experimental data with only gravitational preload, vertical scale 1 N. (b) Numerical simulation of (a), vertical scale 2 N. (c) Displacements of stainless steel and PTFE beads adjacent to the interface. The light green line corresponds to the first PTFE particle just below the interface; the brown line to the first stainless particle above the interface; the red line to the 2$^{nd}$ stainless particle above the interface; the purple line to the 3$^{rd}$ stainless particle above the interface; the light blue line to the 4$^{th}$ stainless particle above the interface; the orange line to the 5$^{th}$ stainless particle above the interface. The displacement of each particle is calculated from the equilibrium positions under zero external field. The slope of the displacement-time curves represents a velocity of the corresponding particle. (d) Diagram showing the relative positions of particles at the moment of 600 μs after the impact. The arrow shows the direction of the displacement $\delta$ .

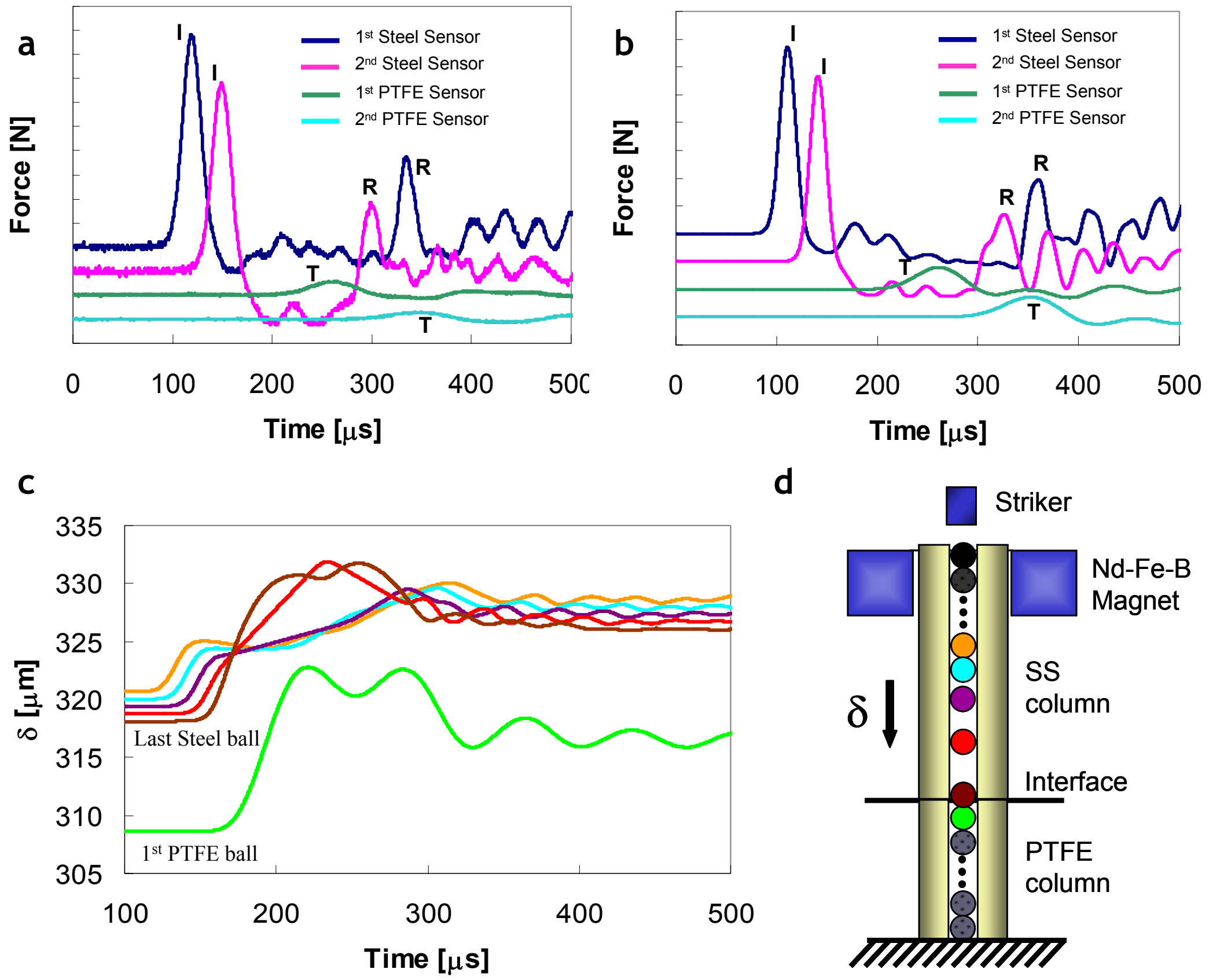

FIG. 2. Anomalous pulse reflection from magnetically preloaded interface of two sonic vacui composed of 20 stainless steel particle and 21 PTFE beads. Letters I, R and T show incident, reflected and transmitted wave transmitted correspondingly. (a) Experimental data for gravitationally and magnetically (2.38 N) precompressed chains, vertical scale 1 N. (b) Numerical simulation of (a), vertical scale 2 N. In (a) and (b) dark blue is an incident wave detected by the sensor in the 8$^{th}$ particle from the interface, pink is the signal from the 4$^{th}$ particle from the interface, green is the signal from the 4$^{th}$ PTFE from the interface, light blue is the signal from the 8$^{th}$ PTFE particle from the interface. (c) Displacements of the stainless steel and PTFE beads adjacent to the interface of the two SVs under magnetic precompression related to the beginning of the formation of the rarefaction wave and anomalous reflected compression waves. The light green is the displacement of the first PTFE particle from the interface; the brown of the first stainless particle from the interface; the red line of the 2$^{nd}$ stainless particle from the interface; the light blue of the 3$^{rd}$ stainless particle from the interface; green line of the 4$^{th}$ stainless particle from the interface; the orange line of the 5$^{th}$ stainless particle from the interface. The displacement of each particle is calculated from equilibrium positions of particles in the chains under zero external field. (d) Diagram showing the relative positions of particles at the moment about 200 microseconds after the impact. The arrow shows the direction of the displacement δ from the particles centres in the initially undeformed state.

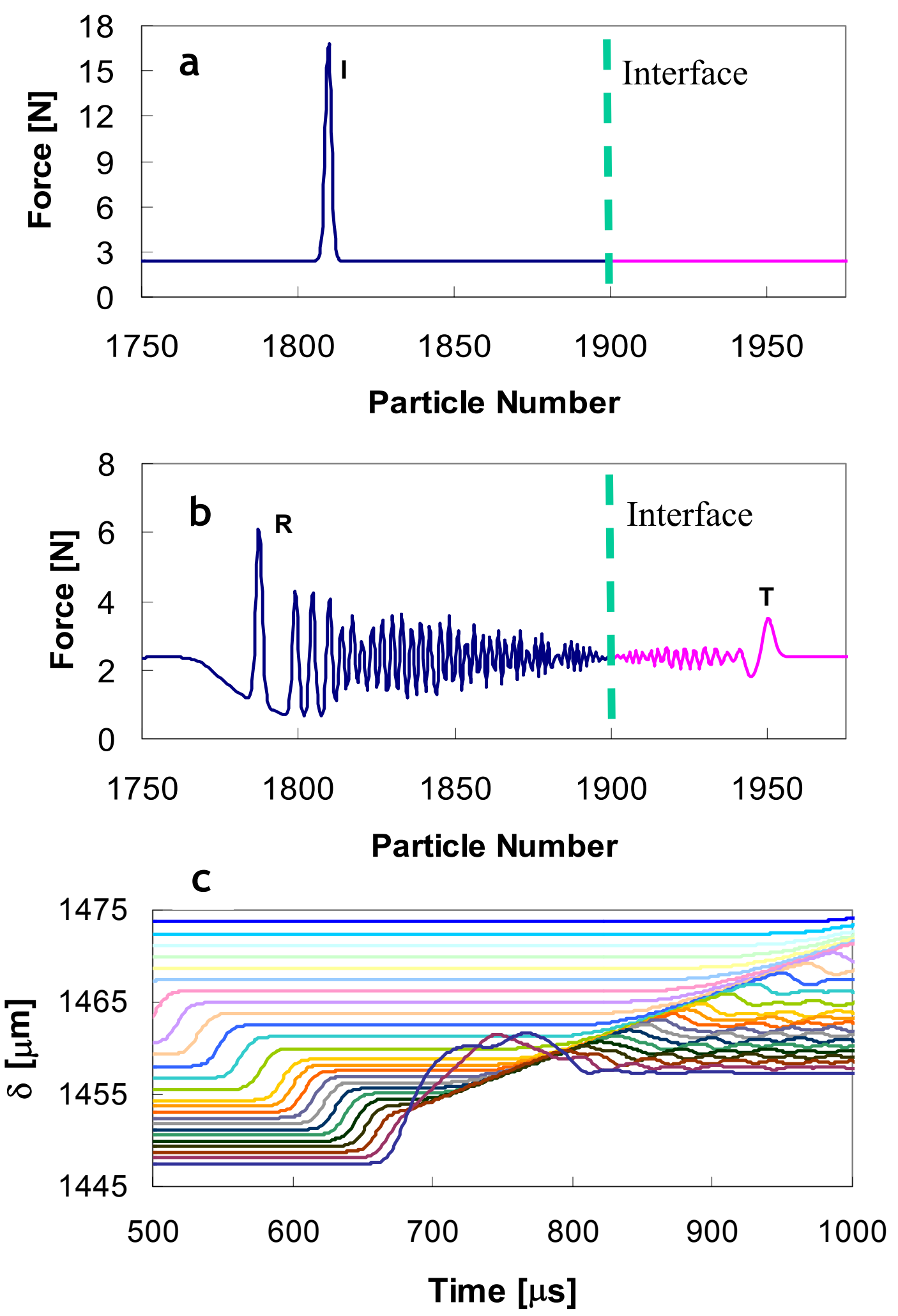

FIG. 3. Solitary pulse reflection from magnetically preloaded interface of two sonic vacui composed of 1900 stainless steel particles and 100 PTFE beads obtained in numerical simulations (no gravitational precompression is included and the magnetic precompression is similar to Fig. 2). The leading incident solitary wave (I) with amplitude similar to Fig. 2 wave was formed after the impact by the alumina striker with the mass equal to mass of the stainless steel particle and a velocity of 0.44 m/s. It was followed by the rarefaction wave which did not participate in the reflection process during the presented time interval. (a) Incident solitary wave (I). (b) Reflected rarefaction wave followed by the anomalous compression waves (R) and an oscillatory tail in the stainless steel chain and transmitted compression (T) (compare with Fig. 1(a) and (b)), rarefaction pulses and oscillatory tail in the PTFE chain. (c) Displacements of the stainless steel beads adjacent to the interface of the two SVs under magnetic precompression related to the beginning of the formation of the reflected rarefaction wave and the anomalous compression waves. The displacement of each particle is calculated from equilibrium positions of particles in the chains under zero external field.